# From river flow to spatial flow: flow map via river extraction algorithm

Zhiwei Wei, Su Ding, Wenjia Xu, Yuanben Zhang, Yang Wang

**Abstract**—Flow maps are thematic maps that visualize object movements across space with a tree layout, in which the underlying tree structure is similar to a natural river system. In this paper, we present a novel and automated approach named RFDA-FM for flow maps from one origin to multiple destinations using a river extraction algorithm in digital elevation models (DEM). The RFDA-FM first models the mapping space as a flat surface by a DEM. A maze-solving algorithm (MSA) for river extraction is then adapted to calculate the flow path from one destination to the origin by constraining its searching directions, direction weights, and searching ranges according to the quality criteria of flow maps. All flow paths from the destinations to the origin are obtained iteratively based on the MSA according to their importance, which is defined by considering their length. Finally, these paths are smoothly rendered with varying widths according to their volume using Bézier curves. A comparison with existing approaches indicates that the flow maps generated by RFDA-FM can be better at keeping nodes away from edges without node overlaps and edge crosses. Two extension experiments demonstrate that RFDA-FM is applicable to heterogeneous mapping space or mapping space with obstacle areas. The parameter analysis shows that RFDA-FM can intuitively control the layouts of flow maps. Project website: https://github.com/TrentonWei/FlowMap

**Keywords:** Flow map; automatic cartography; hydrology model; maze solving algorithm; digital elevation model.

——————————— ◆ ———————————

## 1 INTRODUCTION

Since Henry Drury Harness introduced the first flow map in 1837 [1], it has a long history of depicting migration patterns and good movements with a flow map due to its effective reduction of visual clutter and clear insight into the spatial distribution of moving phenomena [2]. Many approaches such as hierarchical clustering, edge bundling, and force-directed algorithm have been proposed [3-12]. But these works mainly focus on flow maps from multiple origins to multiple destinations. While flow maps from one origin to multiple destinations, i.e., the so-called one-to-many flow maps [2], are also widely used. Pioneers usually use an appealing tree-like layout to delineate the flow routes and volumes which is challenging to produce, thus more attention needs to be drawn to one-to-many flow map production [13].

To automate the production of a one-to-many flow map, some approaches try to generate the tree layout by bundling edges into hierarchical levels, such as the stub bundling(SB), spiral tree (ST), or force-directed (FD) approaches [2, 14-16]. But these approaches may suffer from problems of artificial feeling or low automation. The SB renders all edges as curved ones intentionally where they could be simply straight [13]. The FD requires some manual interventions, for example, the users have to manually define the force factors to obtain a satisfactory layout [15]. Though ST offers automation, it only provides a parameter "the restricting angle" to alter the edge curvature, which is not easy to use for different user demands [13]. The other approaches generate the tree layout by simulating the motion of flows [13, 17]. For example, the river system, as natural flows across geographical regions, can also be considered a special kind of flow. Sun [13] simulates the formation of the dendritic drainage pattern of natural river systems and constructs an approximate Steiner tree to produce the one-to-many flow maps. This approach can produce a natural layout for different mapping spaces, while may generate some redundant nodes and extra operations including edge simplification and edge straightening are required. Inspired by the works of Sun [13, 17], we try to develop an approach that is automatic but intuitive for users to produce a natural one-to-many flow map by simulating the motion of flows.

River extraction is a central research task in hydrological applications, in which the digital elevation model (DEM) is usually used as a basic data model. Many approaches have been proposed in the past decades [18-31] to extract a natural river system automatically. It is intuitive to model the mapping spaces of a one-to-many flow map as DEM data, and then apply these river extraction approaches to automate its production for the following two reasons. First, since DEM is a powerful data model to represent geographic spaces [18], different mapping spaces of flow maps including the heterogeneous mapping spaces or mapping spaces with obstacle areas can be similarly represented by a DEM. Second, existing river extraction approaches such as the maze-solving algorithm (MSA) provide an intuitive way to alter the layout of a river system [25]. To this end, it would be convenient for users to alter the layout of a one-to-many flow map by applying MSA.

Motivated by the above inspirations, we propose a new, automatic approach named RDFA-FM to produce the one-

————————————————

- *Zhiwei Wei, Yang Wang, and Yuanben Zhang are with Aerospace Information Research Institute, Chinese Academic of Sciences and Key Laboratory of Network Information System Technology (NIST), Aerospace Information Research Institute, Chinese Academy of Sciences. E-mails: 2011301130108@whu.edu.cn, Primular@163.com, zhangyb@aircas.ac.cn.*
- *Su Ding is with the College of Environmental and Resource Science, Zhejiang A & F University. E-mail: suding@zafu.edu.cn.*
- *Wenjia Xu is with Beijing University of Posts and Telecommunications. E-mail: xuwen-jia16@mails.ucas.ac.cn.*

to-many flow maps by adapting a river extraction algorithm. We model the mapping space of a flow map as a flat surface with a DEM by defining its grid type, resolution, and range based on user demands[32,33]. Then we adapt the maze-solving algorithm (MSA) to calculate the flow path from one destination to the origin, in which the searching directions, direction weights, and searching ranges of MSA are constrained according to the quality criteria of the one-to-many flow maps. Thus, the involving quality criteria or user demands can all be modeled as the properties of a DEM or the parameters of the river extraction algorithm, which enables the users to be more intuitive to produce a flow map automatically. To further improve the appearance of the flow maps, we smoothly render all flow paths using Bézier curves. Specifically, experiments using different parameter settings on three datasets with homogeneous or heterogeneous mapping spaces or mapping spaces with obstacle areas also demonstrate that the proposed approach is applicable to different mapping spaces and better at keeping nodes away from edges without node overlaps and edge crosses.

The rest of the paper is organized as follows. **Section 2** presents the approaches of the one-to-many flow maps and introduces river extraction algorithms over flat surfaces in a DEM. **Section 3** gives some definitions of the one-to-many flow maps and summarizes the quality criteria. **Section 4** illustrates the methodology of RDFA-FM. **Section 5** evaluates the RDFA-FM, and compares it with previous works. **Section 6** shows some extensions of the RDFA-FM. **Section 7** discusses the strategy effectiveness, parameter settings, and limitations of RDFA-FM. **Section 8** concludes the paper and identifies issues for future works.

## 2 RELATED WORKS

### 2.1 The one-to-many flow maps

The earliest one-to-many flow map can be traced back to Henry Drury Harness in 1837 [1]. Shortly after, Charles Joseph Minard extended this idea to depict economic topics such as the import or export of wine, cotton, and coal [34]. While these early flow maps were most drawn by hand, Tobler [35] first introduced an automatic system to produce flow maps. But straight lines were drawn from destinations to the origin with varying widths in his approach. It will result in visual clutters, which have been the main concern in later approaches.

To reduce the visual clutters, Tobler [35] introduced a filtering strategy. This strategy was also adopted by Elzen and van Wijk [36] in the interaction between flow maps. As an important strategy to reduce visual clutters in graph visualization, edge bundling is also applied for flow maps. Phan et al. [7] proposed an algorithm to bundle edges based on hierarchical clustering. But their approach may not smooth all paths and have a large total graph length. Debiasi et al. [15, 16] used the force-directed algorithm to bundle edges with node merge and node move. But it is a supervised approach and some manual interventions such as suitable force factor settings are required. Steiner tree, as a graph that connects a set of points with minimum total length by using extra points, is also widely used in flow maps [37]. Verbeek et al. [2] introduced a spiral tree by restricting the angle of a Steiner tree to produce a smooth and crossing-free flow map. Nocaj and Ulrik [14] introduced a stub bundling strategy based on a spiral tree. But these approaches don't provide an easy-to-use way to alter the tree layout. Unlike the above approaches, Sun [13, 17] generated the flow maps by simulating the motion of flows according to the formation of the dendritic drainage pattern of natural river systems. The approach can generate a natural layout for different mapping spaces. But his approach may generate some redundant nodes and extra operations including edge simplification and edge straightening are required. Our approach is inspired by Sun [13,17] but with fewer extra operations. Compared to these edge bundling approaches, our approach may have a more natural layout and be more intuitive for users to alter the layout.

### 2.2 River extraction over flat surfaces in a digital elevation model

River extraction in a digital elevation model (DEM) is the key technology in hydrological applications [31]. Its process usually contains four steps: depression filling, flow direction calculation, flow accumulation, and flow track [32]. Because the elevation in a flat surface is the same, depression filling is not performed. Among the remaining three steps, flow direction calculation is the basis for the subsequent two steps. Thus, the flow directions calculation is the key process in river extraction over flat areas in DEMs, which can be categorized into two types: effectiveness first or efficiency first.

Effectiveness first algorithms aim to improve the accuracy of flow direction calculation. Due to the same elevation values within flat areas, Jenson and Domingue [18] assigned flow directions with neighborhood techniques by increasing the elevations from the outlet to the inlet. But this approach may generate parallel flows. Garbrecht and Martz [19] calculated flow directions by adding a gradient from higher to lower terrain. Because the approach of Garbrecht and Martz [19] can produce flow paths that are in line with actual situations, many approaches are then proposed to improve it [20-23]. Another basic idea for flow direction calculation is to obtain flow paths according to their importance iteratively. Tribe [24] extracted flow directions in a flat area by assuming that the main flow path was a straight line from inlet to outlet. Other flow paths were then generated iteratively by connecting them to these earlier generated flow paths. As not all paths are exactly straight lines, Zhang et al. [25] improved the approach by introducing a maze-solving algorithm (MSA) to calculate flow directions. Because we try to generate a one-to-many flow map that meets involving quality criteria as much as possible, the MSA in these effectiveness first algorithms is adopted to calculate the flow paths in our approach.

Efficiency first algorithms aim to reduce the computational cost and mainly involve strategies to reduce or speed up the calculation. For example, Zhu et al. [26] applied a neighbor-grouping scan loop strategy. While Wang and Liu [27] presented a priority-flood algorithm that processed depressions from the edge grid cells to the interior cells. These strategies help reduce the computation complexity. Barnes et al. [28] applied different increments to flat cells in masked DEMs to avoid iterative calculation. Su et al. [29] integrated depression filling and direction assignment with a chain code matrix based on Barnes et al.'s algorithm. Their approach guides the calculation of the gradient directly from higher terrain to lower terrain instead of applying iterative searching. And a distance transform was later introduced to speed up the search process by themselves [30]. Yu et al. [31] used a "first-in, first-out" queue to process depressions and flat areas and a priority queue to process other areas which can effectively reduce the computation complexity.

## 3. Definitions and quality criteria

### 3.1 Definitions

To achieve a better illustration of the proposed approach, some definitions are given. A one-to-many flow map with a tree layout connecting destinations and the origin can be represented as a graph: $G=(V, E)$, where $V$ is the node set representing the destinations and the origin, and $E$ is the edge set connecting two nodes in $V$.

*Origin node*: The node represents the origin, e.g., node $N_1$.

*Destination node*: The node represents a destination, e.g., nodes $N_4$, $N_6$, $N_8$, $N_9$, and $N_{10}$.

*Flow-in node*: The intersection node of two edges, e.g., nodes $N_2$, $N_3$, $N_5$, and $N_7$.

*Hanging edge*: The edge connects a destination node and a flow-in node, e.g., edges $N_4N_2$, $N_6N_5$, $N_8N_5$, $N_9N_7$, and $N_{10}N_7$.

*Non-hanging edge*: The edge which is not a hanging edge, e.g., $N_2N_1$, $N_3N_2$, $N_5N_3$, and $N_7N_3$.

*Path*: The route from one node to another, e.g., $N_{10}N_7N_3N_2N_1$ is a path from the destination node $N_{10}$ to the origin node $N_1$.

*Volume*: The amount of movement objects flowing along an edge, e.g., 100 for edge $N_8N_5$ and 280 for edge $N_7N_3$.

*Flow-in angle*: The angle between an edge and its connecting non-hang edge which is with the largest volume, e.g., angle $N_1N_2N_4$.

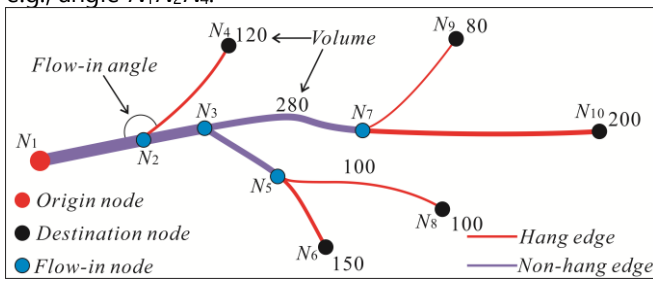

Figure 1. Definitions on the one-to-many flow maps.

### 3.2 Quality criteria

Design requirements of a flow map are usually developed into quality criteria to rule its production, and different quality criteria have been developed for different purposes. For example, Debiasi et al. [15] suggested five criteria including cross-free, minimum overlaps between nodes and edges, etc. Jenny et al. [9] analyzed the design principles for flow maps with a user study and eight kinds of criteria for graph length and symmetry were proposed. Dong et al. [39] analyzed the influence of smoothness and thickness in flow maps using eye-tracking experiments. Xu et al. [40] and Holten and van Wijk [41] analyzed the balance between clarity and aesthetics by using curved edges in graph visualization. To evaluate these quality criteria, measures such as the total length of all edges, the number of edge crosses, the number of acute flow-in angles, visual smoothness index were proposed [13, 42, 43].

We summarize the quality criteria involved in the production of a one-to-many flow map as follows.

(1) **Geometry**

*Curved edge preference* (**$GC_1$**): Curved edges are preferred over straight edges in a flow map, and an edge is more likely to be rendered as a curve if possible [38, 39].

*Curve difference necessity* (**$GC_2$**): The users would prefer curved edges for aesthetics, but prefer straight edges for clarity in graph visualization [40]. Furthermore, the main branches of a flow map also need to be emphasized, and slightly curved edges for edges with high volume are sometimes necessary [15].

*Total length minimization* (**$GC_3$**): Larger total length of a flow map means a larger visual burden, and needs to be minimized [2, 13].

(2) **Relation**

*Acute flow-in angle avoidance* (**$RC_1$**): Acute flow-in angles may have negative impacts on response time, and need to be avoided [44, 45].

*Edge cross avoidance* (**$RC_2$**): Edge crosses may lead to confusion, and need to be avoided [2, 13,15].

*Overlap avoidance between nodes and edges* (**$RC_3$**): Overlaps between destination nodes and edges may lead to visual clutters, and need to be avoided [2, 13, 15].

*Cross avoidance between edges and important map objects* (**$RC_4$**): Some map objects may be important and the edges need to avoid these important map objects to aid recognizability [2, 13].

*Suitable distance between nodes and edges* (**$RC_5$**): The destination nodes should be far away enough from the edges for clarity [13].

(3) **Distribution**

*Tree layout* (**$DC_1$**)*:* The destination nodes should be grouped into hierarchical levels and represented as a tree layout, where the origin is the root node and the destinations are leaf nodes; and edge widths of the flow map are then drawn atop thick ones to the thin ones according to the hierarchy [2, 13, 15].

On the one hand, the quality criteria summarized above may conflict with or enhance one another. For example, though curved edges are preferred($GC_1$), straight lines or slightly curved lines for edges with high volume are also necessary($GC_2$); and keeping a suitable distance between nodes and edges ($RC_5$) will also avoid the overlaps between nodes and edges ($RC_3$). On the other hand, the quality criteria may have different priorities. For example, overlap or cross avoidance ($RC_2$, $RC_3$, and $RC_4$) may have higher priorities than the total length minimization ($GC_3$). Because the overlaps or crosses must be avoided, while the total length minimization only needs to be satisfied as much as possible. Thus, the quality criteria in a one-to-many flow map are not like those in the database domain which should be satisfied completely, they only need to be optimized or fulfilled as much as possible [46].

## 4 METHODOLOGY

### 4.1 Framework

The proposed approach (RFDA-FM) tries to adapt a river extraction algorithm over flat surfaces in a digital elevation model (DEM) to produce the one-to-many flow maps by considering their quality criteria, and RFDA-FM consists of three steps.

**Step 1. Modeling the mapping space as DEM data**: The mapping space is modeled as DEM data over a flat surface by defining its grid type, resolution, and range based on user demands. The destination nodes and the origin node are represented by their corresponding grids, and the destination grids have an outflow that eventually gathers in the origin grid.

**Step 2. Flow path calculation**: The maze-solving algorithm (MSA) is applied to calculate the flow path from one destination grid to the origin grid by constraining its searching directions, direction weights, and searching ranges according to the quality criteria of flow maps. All flow paths from the destination grids to the origin grid are obtained iteratively according to their importance, which is defined by their length.

**Step 3. Flow render**: The edges are rendered smoothly with varying widths according to their flow volumes by using Bézier curves for aesthetic purposes.

### 4.2 Step 1. Modeling the mapping space as DEM data

The digital elevation model (DEM) is a representation of the elevation in a continuous terrain surface via irregular or regular grids with a certain resolution, in which the grid type, resolution, and range need to be defined [42]. Thus, we model the mapping space of a flow map as DEM data by defining its grid type, resolution, and range. Because the mapping space of a flow map is usually considered uniform, the mapping space is modeled as a flat surface, which means the grids are all with the same elevation values.

Suppose the origin node $O$ and destination nodes $N_n$ are represented as $NS = \{O, N_1, ..., N_n, ...\}$, where $N_n$ has a property $f_n$ which represents the flow volume from $N_n$ to $O$. The regions as a base map are $RegS = \{R_1, R_2, ..., R_m, ...\}$, where $R_m$ is a region. The grid type, resolution, and range of the DEM are defined as follows (Figure 2).

(1) *Grids*. Regular grid (square) is adopted by considering its lower computational cost and common uses (Figure 2) [42].

(2) *Resolution*. The resolution needs to make sure that the two closest nodes in the node set ($NS$) do not fall into the same grid. But a higher resolution means a higher computational cost [42]. According to the detailed analysis of resolution setting in spatial applications by Hengl [47], the resolution ($R_s$) of the DEM is defined by considering the shortest distance between nodes in $NS$ as Equation (1).

$$R_s = \frac{AveMinD_{(5\%)}}{4} \qquad (1)$$

Where $AveMinD_{(5\%)}$ is the average distance between the first 5% of the closest node pairs in $NS$.

(3) *Range*. The range of the DEM data can be set according to the envelope of the node set ($NS$) or the regions as a base map ($RegS$), e.g., the range is defined as the envelope of $RegS$ in Figure 2. When the range is defined according to the envelope of $NS$, the range needs to extend a half grid out to prevent nodes in $NS$ from locating on the boundary of the defined range. The two definitions are both available and up to users in practice.

(4) *Modeling the obstacle areas or important objects*. The flow paths sometimes need to avoid obstacle areas or important objects [2,13]. For example, areas with bad weather may need to be avoided while mapping good movements. Thus, grids within these areas need to be deleted from the existing range, e.g., the grids colored red in Figure 2. Based on a similar idea, point, linear, or planar obstacle areas or important objects can all be represented by deleting corresponding grids in our approach (Figure 2). Then cross avoidance between edges and important map objects ($RC_4$) can be satisfied.

(5) *Modeling the heterogeneous mapping space*. Area differences may sometimes be considered by users. For example, bypassing land areas may have a higher travel cost than sea areas while mapping good movements [2, 13]. Two strategies are available and up to users in practice: 1) Different areas are represented as grids with different types; 2) Different areas are represented as grids with different resolutions [31]. As shown in Figure 2, the areas within the regions as a base map ($RegS$) are assigned as Type 2 grids, while the other areas are assigned as Type 1 grids. By defining different types of grids, the areas within $RegS$ or not are distinguished.

With the above definitions, the mapping space is then modeled as DEM data over a flat surface. Each node in the node set ($NS$) can be represented by its corresponding grid as $GS = \{OG, G_1, ..., G_n, ...\}$. Where $OG$ is the grid representing the origin node, and $G_n$ is the grid representing the destination node $N_n$.

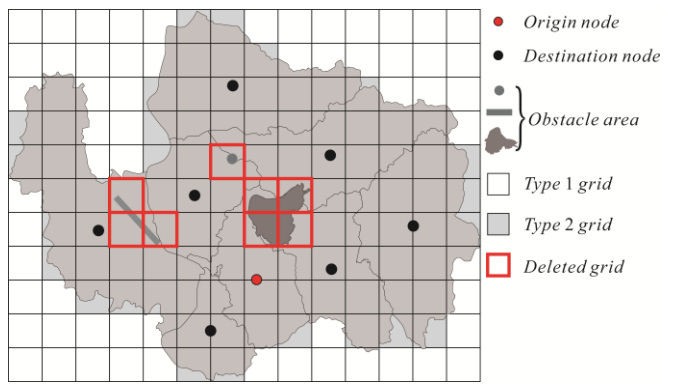

Figure 2. Modeling the mapping space of a flow map as DEM data over a flat surface.

## 4.3 Step 2. Flow path calculation

According to the illustrations in **Section 2.2**, the river extraction calculation in hydrological applications is usually implemented by generating the river paths iteratively according to their importance, in which a less important river path is generated by connecting it to an earlier generated river path. A potential river path from a destination to the origin is defined as a path with the shortest length and is obtained with a maze-solving algorithm (MSA) [24]. Thus, three keys are involved: (1) The path length definition, (2) the MSA, and (3) the path importance definition.

The quality criteria of the one-to-many flow maps need to be optimized or fulfilled as much as possible, as analyzed in **Section 3.2**. Thus, the iterative process for river path calculation is acceptable for flow path calculation in a one-to-many flow map to achieve local optimization. (1) The path length definition, (2) the MSA, and (3) the path importance definition are adapted to calculate the flow paths of a one-to-many flow map by considering its quality criteria as following sections.

### 4.3.1 Path length definition

The length of a path from a destination grid to the origin grid can be computed according to its flowing through grids. Thus, the length between two neighborhood grids is defined first. The length of a path is then defined by considering the quality criteria of the one-to-many flow maps.

#### 4.3.1.1 *Length definition between two neighborhood grids*

The grids may have different types if the mapping space is heterogeneous, and can be assigned different weights (as defined in **Section 4.2**). The length ($NL$) between two neighborhood grids ($G_m$ and $G_n$) is defined as Equation (2), as shown in Figure 3.

$$NL = \begin{cases} 0.5R_s \times \delta_m + 0.5R_s \times \delta_n & \text{(Orthogonal neighbors)} \\ \sqrt{2} \times (0.5R_s \times \delta_m + 0.5R_s \times \delta_n) & \text{(Diagonal neighbors)} \end{cases} \quad (2)$$

Where $R_s$ is the resolution of the DEM data (defined in **Section 4.2**), $\delta_m$ and $\delta_n$ are the weights of $G_m$ and $G_n$ and can be set by users. If the mapping space is considered homogeneous, grids are all the same type and $\delta_m = \delta_n = 1$.

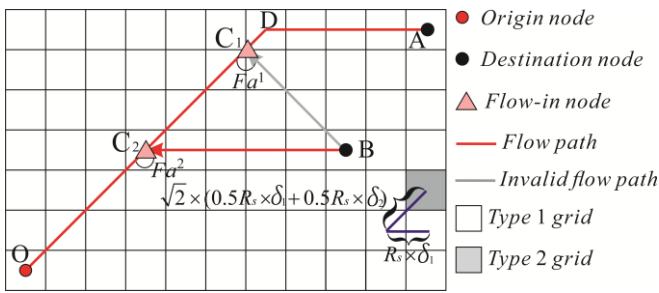

Figure 3. Path length definition. If only the total length minimization requirement ($GC_3$) is considered, the path $BC_1O$ is preferred over $BC_2O$ because $C_1O$ and $C_2O$ are not newly generated sub-paths, and the newly generated sub-path $BC_1$ is shorter than $BC_2$. If the acute flow-in angle avoidance ($RC_1$) is also considered, the path $BC_2O$ is preferred over $BC_1O$ because the length of the path $BC_1O$ is penalized for its acute flow-in angle.

#### 4.3.1.2 *Path length definition and refinement*

The paths from the destination grids to the origin grid are generated iteratively in our approach. Suppose the earlier generated paths are $FpS = \{Fp_1, Fp_2, ..., Fp_n, ...\}$, $Fp_n$ is an earlier generated path from a destination grid ($G_n$) to the origin grid ($OG$). For a newly given destination grid ($G_i$), a potential path ($Fp_i$) from $G_i$ may flow in any grid of $Fp_n \in FpS$ to $OG$. Because the valid flow path from $G_i$ to $OG$ is defined as a path with the shortest length in hydrological applications [24]. Thus, when we try to obtain the valid path from $G_i$ to $OG$, we need to find all the shortest paths ($CacheFpS_i$) from $G_i$ that flow in each grid of $Fp_n \in FpS$ with the maze-solving algorithm (MSA) first. And the valid path from $G_i$ to $OG$ should be the shortest one in $CacheFpS_i$. Thus, if $Fp_i$ violates a quality criterion in **Section 3.2**, it will have a larger length and be less likely to be the valid path from $G_i$ to $OG$ and vice versa. Two quality criteria, total length minimization ($GC_3$) and acute flow-in angle avoidance ($RC_1$), are considered here for path length definition.

**(1) Path length definition by considering the total length minimization ($GC_3$)**

For a potential path ($Fp_i$) from a destination grid ($G_i$) to the origin grid ($OG$), $Fp_i$ can be divided into two parts: $sub\text{-}Fp^1_i$ and $sub\text{-}Fp^2_i$, where $sub\text{-}Fp^2_i$ is a sub-path of $Fp_n \in FpS$ and is not newly generated, and $sub\text{-}Fp^1_i$ is newly generated. Then a shorter $sub\text{-}Fp^1_i$ will help minimize the total length ($GC_3$). For example, the two paths ($BC_1O$ and $BC_2O$) in Figure 3 that separately flow in an earlier generated path ($ADO$) at grid $C_1$ and $C_2$ can both be a path from the destination grid B to the origin grid O. Because $C_1O$ and $C_2O$ are both parts of an earlier generated path ($ADO$), but the sub-path $BC_1$ of $BC_1O$ is shorter than the sub-path $BC_2$ of $BC_2O$. If the total length minimization requirement ($GC_3$) is considered, $BC_1O$ is more likely to be the valid path from destination grid B to origin grid O due to a smaller total length. This means that the length of $sub\text{-}Fp^1$ should be more important than the length of $sub\text{-}Fp^2_i$ in the path length definition if $GC_3$ is considered. Then the length ($PL_i$) for $Fp_i$ is defined as Equation (3).

$$PL_i = subPL^1_i + subPL^2_i \times \omega \quad (3)$$

Where $subPL^1_i$ and $subPL^2_i$ are the lengths of $sub\text{-}Fp^1_i$ and $sub\text{-}Fp^2_i$, $\omega$ is a weight and $\omega \leq 1$, it means $subPL^1_i$ is more important than $subPL^2_i$.

**(2) Path length refinement by considering the acute flow-in angle avoidance (RC1)**

The acute flow-in angle avoidance ($RC_1$) rules that acute flow-in angles ($Fa$, defined in **Section 3.1**) need to be avoided. Whether a $Fa$ is an acute one is defined by setting a threshold ($T_a$): If $Fa \leq T_a$, then $Fa$ is an acute one. If a potential path ($Fp_i$) from a destination grid ($G_i$) to the origin grid ($OG$) flows in an earlier generated path with an acute flow-in angle, a penalty strategy is applied to refine its length. Thus, $Fp_i$ is less likely to be the shortest (or valid) one from $G_i$ to $OG$. The path length refinement is defined as Equation (4).

$$PL_i = PL_i + PL_{pen} \quad (4)$$

Where $PL_i$ is the length of $Fp_i$, $PL_{pen}$ is a large constant and set as $PL_{pen} = 20R_s$ in this approach. For example, the two paths $BC_1O$ and $BC_2O$ in Figure 3 can be both a path from the destination grid B to the origin grid O. But the path $BC_1O$ flows in an earlier generated flow path ($ADO$) at grid $C_1$ with an acute flow-in angle ($Fa^1$). Then the path length of $BC_1O$ is penalized by Equation (4), and $BC_2O$ is finally obtained as the shortest path from destination grid B to origin grid O because its flow-in angle ($Fa^2$) at grid $C_2$ is not an acute one.

### 4.3.2 The maze-solving algorithm

Maze solving is a classical problem in graph theory and data structure fields. As it aims to find the shortest path between the entrance to the outlet in a given labyrinth, which has been applied to generate river paths in digital elevation model (DEM) data successfully [25]. The flat surface is considered as a labyrinth without inside walls, the destination grid is considered an entrance, and the origin grid is considered an outlet. Three definitions are required in a maze-solving algorithm (MSA): (1) searching directions, (2) direction weights, and (3) searching ranges [48]. Searching directions determine the potential directions to be searched. Direction weights determine which direction is first to be searched. Searching ranges determine where can be searched. They are defined by considering the quality criteria in **Section 3.2**.

#### 4.3.2.1 Searching direction definition

Searching directions determine the potential directions to be searched. As a grid in DEM, its outflow direction to its neighborhood grids can be 8 directions (8D), denoted as $8D = \{0,1,2,3,4,5,6,7\}$ [49]. However, it is time-consuming if all directions are explored each time and it might be more suitable to only search in the directions towards the target grid. As shown in Figure 4, the searching directions might be $\{0,1,2\}$ while searching from grid A to grid N, and $\{2,3,4\}$ while searching from grid B to grid N.

Given a start grid as $G_i$, the target grid that $G_i$ searches towards is $G_j$, and the angle between the line $G_iG_j$ which connects $G_i$ and $G_j$ and the horizontal direction is $Angle_{ij}$. The searching directions ($sDir_{ij}$) from $G_i$ to $G_j$ are defined as Equation (5).

$$sDir_{ij} = \{x | x = y \bmod 8, y \in \{z-1, z, z+1\}, z = \lfloor Angle_{ij} / 45 \rfloor \} \quad (5)$$

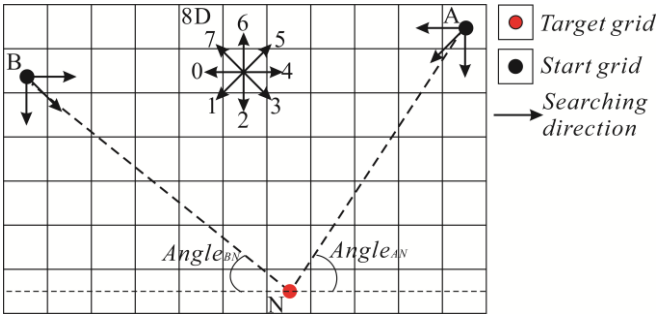

Figure 4. Searching direction definition. It only searches in the directions from the start grid toward the target grid to improve efficiency.

#### 4.3.2.2 Direction weight definition

Direction weights determine which direction is first to be searched. Direction weights in digital elevation models (DEM) are usually assigned as the downward slope angle of neighborhood grids, which are defined according to the distance and elevation value difference between two neighborhood grids [19]. As the mapping space is modeled as a flat surface in this approach, and grids are all with the same elevation values. Then the elevation value difference between the two neighborhood grids is 0. To overcome this problem, the idea according to river system extraction in DEM data that the grid with higher flow accumulation is more likely to be mainstream is adopted. This idea can also help generate a more suitable tree layout ($DC_1$) and minimize the total length ($GC_3$)[25].

The grid with higher flow accumulation is more likely to be mainstream, which means the direction toward a grid that has a larger potential flow accumulation is prior to be explored. The potential flow accumulation ($Pf_i$) of a grid ($G_i$) can be calculated by the total volume of the grids in its $k$-order surrounding and is defined as Equation (6) [25].

$$Pf_i = \sum_{x=m-k, y=n-k}^{x \le m+k, y \le n+k} f_{xy} \quad (6)$$

Where $G_i$ is a grid in row $m$ and column $n$ in the DEM data, $f_{xy}$ is the volume of the grid in row $x$ and column $y$. Given two neighborhood grids as $G_i$ and $G_j$, the direction weight ($DW_{ij}$) for the direction from $G_i$ to $G_j$ is defined as Equation (7) [19, 25]

$$DW_{ij} = \begin{cases} (Pf_i - Pf_j + Tf) / NL_{ij} & (Pf_i - Pf_j > 0) \\ Tf / NL_{ij} & (Pf_i - Pf_j \le 0) \end{cases} \quad (7)$$

Where $Pf_i$ and $Pf_j$ are the potential flow accumulation of $G_i$ and $G_j$, $Tf$ is a constant, and $Tf \in R^+$ in case $Pf_i - Pf_j \le 0$, $NL_{ij}$ is the length between $G_i$ and $G_j$ (defined as Equation (2)).

As shown in Figure 5, the two paths (AEDO and ACFO) may both be as a path from destination grid A to the origin grid O if the potential flow accumulation ($Pf$) of each grid is not considered. But the path ACDO is finally obtained as a path from destination grid A to the origin grid O by considering the $Pf$ of each grid. Though ACDO is of the same length as AEDO and ACFO, it is clear in Figure 5 that more destination grids are more likely to connect ACDO with shorter paths than to connect AEDO and ACFO. Thus, the direction weight definition with Equation (7) will help generate a more suitable tree layout ($DC_1$) and minimize the total length ($GC_3$).

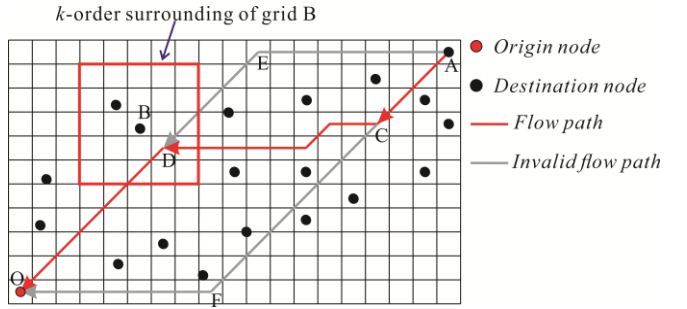

Figure 5. Direction weight definition. The searching direction toward a grid that has a larger potential flow accumulation is prior to be explored.

#### 4.3.2.3 Searching range definition

The maze-solving algorithm (MSA) is performed with defined directions and direction weights in a searching range to obtain the shortest path between two grids. If certain grids are excluded from the searching range, then the path obtained between two grids by MSA will never flow through the certain grids, namely, no overlaps or crosses will be made on these grids. As shown in Figure 6, if we exclude grid B in the searching range while we try to obtain a path from the destination grid A to the origin grid O with the MSA, then all the potential paths from the destination grid A to the origin grid O will never overlap grid B.

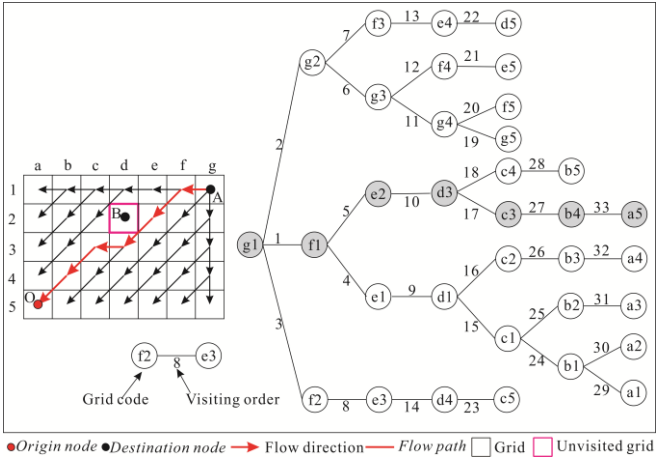

Figure 6. An example of the shortest path calculation with the maze-solving algorithm(MAS) in a defined searching range. Grid B is excluded from the searching range, and all the potential paths from grid A to grid O will never cross or overlap grid B.

The quality criteria in **Section 3.2** rule that overlaps between nodes and edges should be avoided ($RC_3$) and suitable distance between nodes and edges should be maintained ($RC_5$). As the example shown in Figure 6, if we exclude the grids representing target nodes in the searching range, the obtained paths with the maze-solving algorithm (MSA) will never overlap the target nodes. Because the paths may be rendered with varying widths and suitable distances also need to be maintained between nodes and edges. We can exclude the $t$-order surrounding of specific grids from the searching range to satisfy $RC_3$ and $RC_5$, where two situations need to be specifically considered.

**(1) Special consideration for $t=1$**

The destination nodes may not be exactly in the center of their representing grids, e.g., the destination node in grid B in Figure 7(a). Because the destination node in grid B locates near the boundary of its representing grid, the distance between the obtained path ACO and the node in grid B may be small though the path does not overlap grid B with $t=1$. To maintain suitable distances between nodes and edges with $t=1$, we define 1-order surrounding ($1\text{-}SG_i$) for a grid ($G_i$) representing a node ($N_i$) as Equation (8).

$$1\text{-}SG_i = \{G_{xy} | m-1 \leq x \leq m+1, n-1 \leq y \leq n+1,$$
$$HDis(G_{xy}, N_i) < 0.5R_s, VDis(G_{xy}, N_i) < 0.5R_s\} \quad (8)$$

Where $G_i$ is a grid in row $m$ and column $n$ in the digital elevation model (DEM) data, $G_{xy}$ is a grid in row $x$ and column $y$, $R_s$ is the resolution of the DEM data (defined as equation (1)). $HDis(G_{xy}, N_i)$ and $VDis(G_{xy}, N_i)$ are the minimum distances between grid $G_{xy}$ and node $N_i$ in horizontal and vertical directions. The defined 1-order surrounding for grid B is shown in Figure 7(a).

**(2) Special consideration for particularly close destination nodes**

If the destination nodes are close enough, then a given destination grid ($G_i$) may be in the $t$-order surrounding of another grid. Then $G_i$ will be excluded from the searching range according to our definition, and no available paths can be obtained from $G_i$ to the origin grid ($OG$). Thus, smaller $t$ is adaptively set for these grids in our approach to avoid $G_i$ being excluded from the searching range. As shown in Figure 7(b), grid A is in a 2-order surrounding of grid B. If we set $t=2$ while we try to obtain a path from grid A to the grid O with the maze-solving algorithm(MAS), grid A will be excluded from the searching range and no available paths can be obtained from grid A to grid O. Thus, $t=1$ is adaptively set for grid B on this occasion.

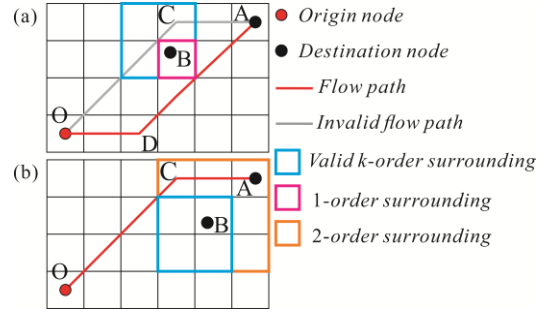

Figure 7. Special considerations for searching range definition. (a) Special consideration for $t=1$: the grids colored cyan are the defined 1-order surrounding of grid B; (b) Special consideration for close destination nodes: $t$ is adaptively set from $t=2$ to $t=1$ for grid B because grid A is in 2-order surrounding of grid B.

### 4.3.3 Path importance

River importance is usually determined by its length and flow volume [25]. As ruled by the total length minimization ($GC_3$), the path length may be the first to be considered. Thus, the importance($FP_{im}$) of a flow path ($Fp_i$) is defined by considering its length in our approach. Because the paths are generated iteratively in this approach, supposed the earlier generated flow paths are $FpS = \{Fp_1, Fp_2, ..., Fp_n, ...\}$, $Fp_n$ is a path from a destination grid ($G_n$) to the origin grid ($OG$). For a newly generated path ($Fp_i$) from a destination grid ($G_i$) to $OG$, $Fp_i$ can be classified into two types by considering whether its sub-path is part of $Fp_n \in FpS$, as follows.

**Type I flow path**: if no sub-path of $Fp_i$ is a part of $Fp_n \in FpS$, then $Fp_i$ is a type I path.

**Type II flow path**: If there exists a sub-path of $Fp_i$ is a part of $Fp_n \in FpS$, then $Fp_i$ is a type II path.

If $Fp_i$ is a type I path, it means that $Fp_i$ from the destination grid straightly connects to the origin grid. Then $Fp_i$ is more likely to be another mainstream according to the river system extraction process in hydrological applications [25]. Thus, the type I path is considered more important than a type II path, and the path importance ($FP_{im}$) of $Fp_i$ is defined as Equation (8).

$$FP_{im} = \begin{cases} PL_i + PL_{im} & (\text{Type I } Path) \\ PL_i & (\text{Type II } Path) \end{cases} \quad (8)$$

Where $PL_i$ is the length of $Fp_i$, and $PL_{im}$ is a large constant to make sure that a type I path is more important than a type II path, and $PL_{im} = 10000R_s$ in this approach.

### 4.3.4 Iterative process for flow path calculation

Given the origin grid ($OG$) and destination grids ($G_n$) as $GS = \{G_1, ..., G_n, ...\}$ in modeling DEM data according to **Section 4.2**, the flow paths from all destination grids to the origin grid are generated iteratively as **Algorithm 1**.

Because the flow path from a destination grid to the origin grid is defined as a path with the shortest length, and the path with the largest importance is selected iteratively. Thus, the iterative process can naturally avoid edge crosses ($RC_2$) according to the river extraction process [18, 19].

---

**Algorithm 1: Flow path calculation**

**Input**: The origin grid is $OG$, destination grids are $GS = \{G_1, ..., G_n, ...\}$, and the DEM data represents the mapping space.

**Output**: Flow paths from $G_n \in GS$ to $OG$ as $FpS = \{Fp_1, Fp_2, ..., Fp_n, ...\}$

**Initialization**: The output flow paths as $FpS = Null$, the grids representing the paths in $FpS$ as $FGS = Null$, and the two cache flow paths as $CacheFpS_1 = Null$ and $CacheFpS_2 = Null$, $CacheFpS_1$ records the all shortest paths from each destination grid to $OG$, $CacheFpS_2$ records the all potential path from a given destination grid to $OG$.

**While** $GS$ **Not** $Null$ **Do**:
  Set $CacheFpS_1 = Null$;
  **Foreach** $G_n \in GS$ **Do**:
    Set $CacheFpS_2 = Null$;
    **Foreach** $G_m \in FGS$ **Do**:
      Get flow path ($Fp_i$) with the shortest '**path length**' from $G_n$ to $OG$ which flows in a path in $FpS$ at grid $G_m$ with '**the maze solving algorithm**'; add $Fp_i$ to $CacheFpS_2$;
    Get the path ($Fp_j$) with the shortest '**path length**' in $CacheFpS_2$ and add it to $CacheFpS_1$;
  Get the path ($Fp_L$) with the largest '**path importance**' in $CacheFpS_1$ and add it to $FpS$; Update the grids in $FGS$; Remove the corresponding destination grid of $Fp_L$ from $GS$
**Return** $FpS$

---

## 4.4 Step 3. Flow render

After all flow paths are obtained, we render the flow paths with varying widths according to their volumes. To further improve their appearance, a smooth process is implemented to render these flow paths as curves.

### 4.4.1 The edge widths

The width of an edge is set as the common practice in other works [2, 13-17], being proportional to its volume which can be obtained based on the flow accumulation of its flowing through grids. As shown in Figure 8(a), the flow accumulation of a certain grid is the total amount of the flow volume which gathers at the grid and can be obtained based on the calculated flow paths. The result to render the flow paths with varying widths is shown in Figure 8(b).

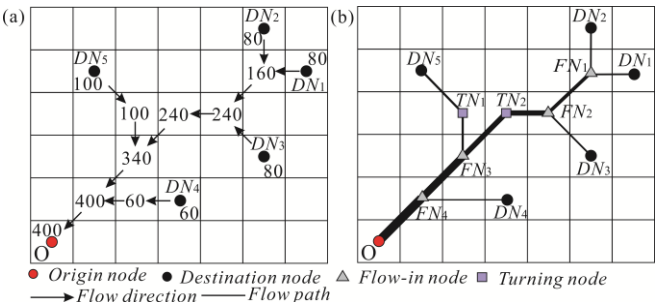

Figure 8. The edge widths. (a) Flow accumulation of each grid; (b) The edges are rendered with their widths being proportional to their flow volume.

### 4.4.2 Smooth process

We use the Bézier curve to smooth the edges according to Sun [13] by considering the curved edge preference ($GC_1$) and curve difference necessity ($GC_2$). Because the edges are rendered with varying widths, the flow-in locations may become not smooth, e.g., the locations at flow-in nodes $FN_1$, $FN_2$, $FN_3$, and $FN_4$ in Figure 8(b). Thus, the nodes at these flow-in locations are shifted first and smooth operations are then implemented.

(1) **Node shifts at the flow-in locations**

To enable smoothness at the flow-in locations, the nodes of a branch edge connecting its main branch are shifted with a distance in a direction that is perpendicular to the vector of its connecting main branch [15, 16]. The shifted distance ($Sd$) is decided by all edges involving the flow-in locations. Given a flow-in node as $N_n$, the edges which flow in $N_n$ are arranged in a clockwise order as $ES = \{E_1, E_2, ...E_i, ...\}$, the width of $E_i$ is $W_i$ and the width of the main branch is $W_{main}$. The shifted distance ($Sd$) of the flow-in node for a given edge ($E_i$) is defined as Equation (9) (Figure 9).

$$Sd = \sum_{j=0}^{j \leq i-1} W_j + 0.5*W_i - 0.5*W_{main} \qquad (9)$$

Here the shifted distance ($Sd$) is graphic and is computed based on the widths of the edges. Because flow maps are mainly applied to visualize object movements across geographic spaces. If scale or projection is considered in a flow map, $Sd$ needs to be converted into geographic distance. We have offered a tool to convert graphic distance to geographic distance in our uploaded code.

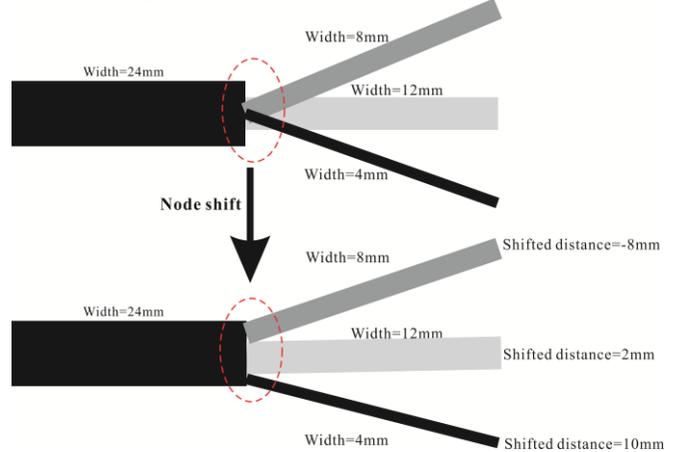

Figure 9. Node shifts at the flow-in locations.

(2) **Smooth the edges with Bézier curves**

We smooth the edges with Bézier curves by mainly considering the curved edge preference ($GC_1$) and curve difference necessity ($GC_2$). $GC_1$ rules that an edge is more likely to be rendered as a curve if possible. While $GC_2$ rules that the main branches of the flow map need to be emphasized, slightly curved lines for edges with high volume are also necessary. Here the non-hanging edges can all be considered as the main branches. Thus, the hanging edges are smoothed as regularly curved lines, and the non-hanging edges are smoothed as slightly curved lines. Furthermore, because the edges are rendered with varying widths, the smooth process also needs to enable smoothness at the flow-in locations.

A Bézier curve is drawn by defining its controlling nodes [13]. The start node, turning nodes and end node of a given edge ($E_i$) can be considered as the initial controlling nodes of the Bézier curve, and two new controlling nodes need to be inserted for $E_i$ to create curved lines. Suppose the controlling nodes of $E_i$ are $CNS = \{N_1, N_2, ..., N_n..., N_e\}$, where $N_1$ is the start node of $E_i$, $N_n$ is a turning node of $E_i$, $N_e$ is the end node of $E_i$, and the vector of $E_i$'s connecting main branch is $\vec{vec_t}$. Two new controlling nodes are inserted as follows.

**The first inserted controlling node ($N_p$):** $N_p$ is inserted for $E_i$ to enable smoothness at the flow-in locations. $N_p$ is inserted by making the distance of $N_1N_p$ ($Dis_{1p}$) as $Dis_{1p} = setDis$, $N_1N_p$ is in direction of $\vec{vec_t}$, as shown in Figure 10(b). $setDis$ is a small constant and is set as $setDis = 0.2R_s$ in our approach ($R_s$ is the resolution of the DEM data).

**The second inserted controlling node ($N_q$):** $N_q$ is inserted to control the curvature level of the curve by considering whether the given edge ($E_i$) is a hanging edge or not: If $E_i$ is a hanging edge, $N_q$ is added by making the distance of $N_1N_q$ ($Dis_{1q}$) as $Dis_{1q} = \alpha * Dis_{12}$, $n_1n_q$ is in direction of $\vec{vec_t}$; If $E_i$ is a non-hanging edge, $N_q$ is added by making the distance of $n_1n_q$ ($Dis_{1q}$) as $Dis_{1q} = \beta * Dis_{12}$, $n_1n_q$ is in direction of $\vec{n_1n_2}$, as shown in Figure 10(b). Where $\alpha$ and $\beta$ are constants, $Dis_{12}$ is the length of $N_1N_2$, $\alpha=0.5$ to make sure the rendered line is a regular curved one, and $\beta=0.1$ to make sure the rendered line is a slightly curved one, $\alpha$ and $\beta$ can also be set by users to control the edge curvature. Because the given edge ($E_i$) may have turning nodes, and three nodes are enough to draw a Bézier curve. If the number of the initial controlling nodes of $E_i$ is not less than 3, we won't insert a second controlling node for $E_i$, e.g., the second controlling nodes are not inserted for edges $FN_3TN_2FN_2$ and $FN_3TN_1DN_5$ in Figure 10 (b).

## 5 EXPERIMENTS

### 5.1 Experimental settings

#### (1) Datasets

In this paper, we take the population migrations from Texas to other states in the 2000 US Census as experiment data, which has been a widely used benchmark dataset for flow map production [2, 13-17], and comparisons are then made. Data source as: https://www.ipums.org/.

Two other datasets are also taken as experimental data in extensions of our approach, in which mapping spaces with obstacle areas and heterogeneous mapping spaces are considered. They will be illustrated in **Section 6**.

#### (2) Parameter settings

Three parameters need to be set in our approach.

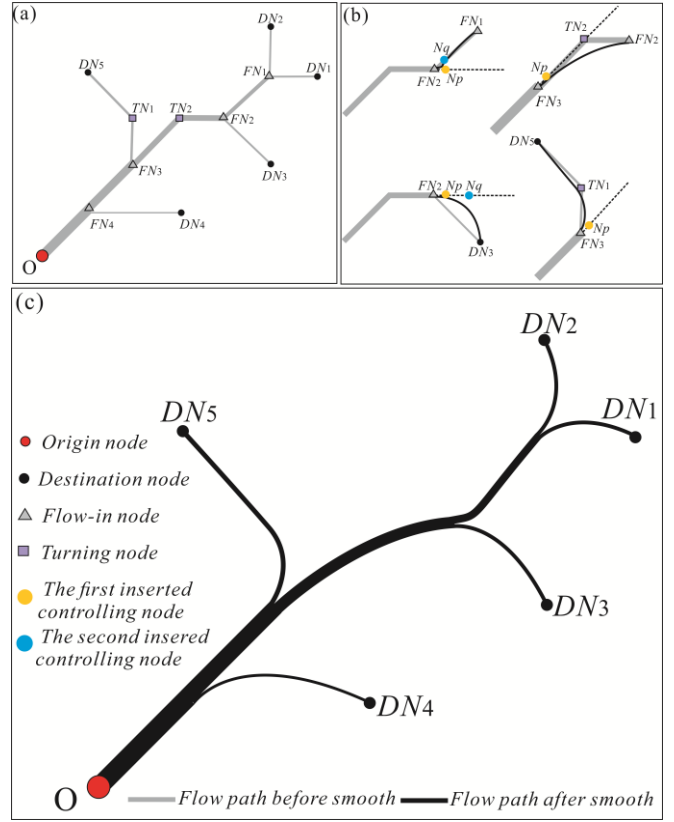

Figure 10. Flow smooth. (a) The flow paths before smooth; (b) Inserted controlling nodes; (d) The flow paths after smooth.

The parameters to generate the flow map for evaluations are set according to experiments as follows: $\omega=0.65$ in Equation (3), $k=4$ in Equation (6), and $t=1$ in searching range definition by considering overlap avoidance and suitable distance maintenance between nodes and edges ($RC_3$ and $RC_5$).

The influence of the parameter settings on flow map layouts will be discussed in detail in **Section 7.2**. Two other layouts with different parameter settings will be displayed in comparison with other approaches, and more layouts with different parameter settings will be displayed in **Sections 7.1 and 7.2**.

#### (3) Evaluation metrics

The resulting quality is evaluated according to the quality criteria in **Section 3.2**. The metrics are adopted as follows [13, 42, 43].

The mean value of visual smoothness index (*MSI*): Visual smoothness index (*SI*) is a metric which is proposed by Sun [13]. *SI* gives high values to the smooth edges with small splitting angles and punishes unnecessary bends and large curvature. Therefore, visually smooth and natural edges will have high values. The mean value of *SI* (*MSI*) for all edges is adopted to evaluate the curved edge preference ($GC_1$) and curve difference necessity ($GC_2$).

The total length of the flow map (*TL*): *TL* is used to evaluate the total length minimization ($GC_3$).

The number of acute flow-in angles (*FaN*): *FaN* is used to evaluate the acute flow-in angle avoidance ($RC_1$). Acute flow-in angle(*Fa*) is defined by setting a threshold ($T_a$) for *Fa*. If $Fa \leq T_a$, then *Fa* is an acute one, and $T_a=120$ in our approach.

The number of edge crosses (*ECN*), the number of overlaps between nodes and edges (*EON*), and the number of crosses between edges and important objects (*EIN*): *ECN*, *EON*, and *EIN* are used to evaluate the overlap or cross avoidance (*RC*$_2$, *RC*$_3$, and *RC*$_4$).

The minimum distance between nodes and their nearby edges (*MDis*): *MDis* is used to evaluate the suitable distance maintenance between nodes and edges (*RC*$_5$) according to Sun[13].

The mean value of user preference index (*MUPI*): user preference index (*UPI*) is a preference value rated by users for a flow map layout. The mean value of *UPI* is used to evaluate whether a flow map layout is preferred by users (*DC*$_1$). *UPI* is obtained by asking 23 participants to rate how much they liked the layout with a 5-point scale. 1 means "very dislike", 2 means "dislike", 3 means "Neither dislike nor like", 4 means "like", and 5 means "very like". The participants are postgraduate or undergraduate students from Wuhan University, and they all major in cartography or geographic information system.

## 5.2 Results and evaluations

The generated flow maps (RFDA-FM$_1$) for population migration from Texas to other states in the 2000 US Census with given parameter settings are shown in Figure 11. The statistical results are shown in Table 1. From Figure 11 and Table 1, we have the following observations: (1) All edges are rendered with varying widths according to their volumes, and they are smoothed by using Bézier curves and *MSI* is 0.751 (*GC*$_1$ and *GC*$_2$ are satisfied). (2) No acute flow-in angles are made (*RC*$_1$ is satisfied); (3) No edge crosses are made (*RC*$_2$ is satisfied); (4) No overlaps between nodes and edges are made (*RC*$_3$ is satisfied); (5) The minimum value of *MDis* is 47.6*10$^3$m and 14 nodes are with their *MDis* less than 100*10$^3$m, these mean that suitable distances can also be maintained between nodes and edges (*RC*$_5$ is satisfied); (6) *MUPI* is 3.30 out of 5 (5 being the optimal value), which mean that most of the users moderately prefer the generated flow map (*DC*$_1$ is satisfied). With the above observations, we can conclude that the flow map generated with our approach can well meet all the quality criteria in **Section 3.2**.

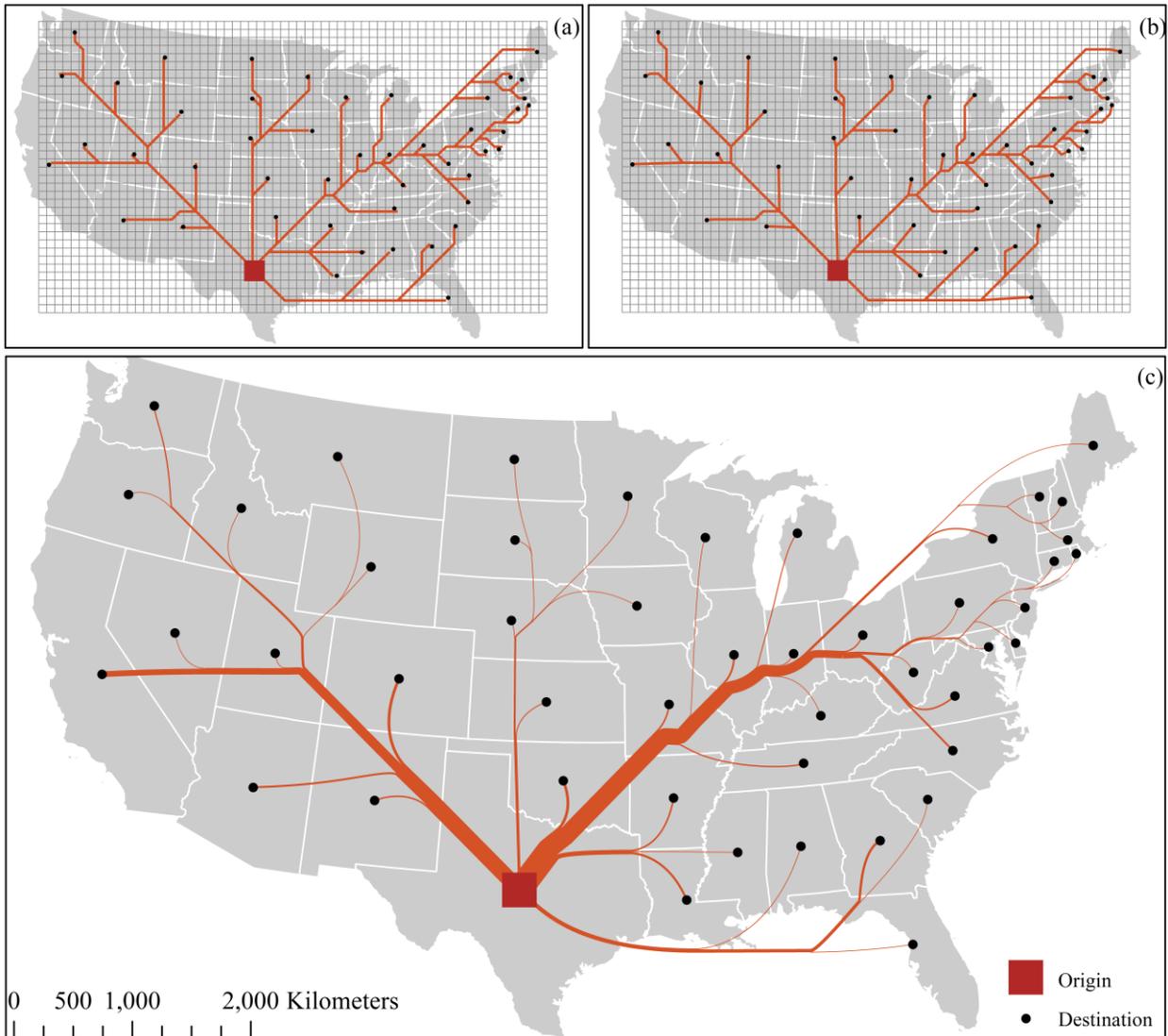

Figure 11. Flow map for population migration from Texas to other states in 2000 US Census. (a) Flow directions in the digital elevation model (DEM); (b) Flow paths in which the nodes are connected based on the calculated flow directions; (c) Flow map in which the flow paths are smoothed and rendered with varying widths (RFDA-FM$_1$).

Table 1. Quantitative assessment on layout quality.

| Measures | RFDA-FM$_1$ | RFDA-FM$_2$ | RFDA-FM$_3$ | FD | ST | SB | TNSS |
|---|---|---|---|---|---|---|---|
| $TL$ ($10^6$m) ↓ | 24.63 | 22.58 | 25.03 | 23.61 | 22.01 | 24.46 | 25.59 |
| $MSI$ ↑ | 0.751 | 0.758 | 0.769 | 0.625 | 0.621 | 0.698 | 0.726 |
| $MDis$ ($10^3$ m) | | | | | | | |
| $MDis_{min}$ ↑ | 47.6 | 52.5 | 38.9 | 7.8 | 49.7 | 13.0 | 37.7 |
| $n$ ($MDis$<100) ↓ | 15 | 17 | 6 | 29 | 27 | 28 | 15 |
| $n$ ($MDis$ <70) ↓ | 8 | 8 | 4 | 27 | 14 | 17 | 2 |
| $n$ ($MDis$ <40) ↓ | 0 | 0 | 1 | 5 | 0 | 10 | 0 |
| $n$ ($MDis$ <20) ↓ | 0 | 0 | 0 | 1 | 0 | 1 | 0 |
| $FaN$ (<120°) ↓ | 0 | 0 | 0 | 1 | 0 | 0 | 4 |
| $ECN$ ↓ | 0 | 0 | 0 | 0 | 0 | 0 | 0 |
| $EON$ ↓ | 0 | 0 | 0 | 1 | 0 | 0 | 0 |
| $MUPI$ ↑ | 3.30 | 3.04 | 3.43 | 3.13 | 3.26 | 3.52 | 3.39 |

**Note**: RFDA-FM is the new approach proposed in this article, where RFDA-FM$_1$ is Figure 11(c), RFDA-FM$_2$ is Figure 12(e), and RFDA-FM$_3$ is Figure 12(f); FD is the force-directed approach [15], Figure 12(a); SB is the stub bundling approach [14], Figure 12(b); ST is the spiral tree approach [2], Figure 12(c); TNSS is the flow motion simulation approach [13], Figure 12(d). Some statistics in Table 1 for TNSS, ST, SB, and FD are from Sun [13].

## 5.3 Comparisons

To validate the feasibility and generalization ability of the proposed approach (RFDA-FM), we compared the obtained flow maps with the proposed approach to existing approaches, including the force-directed approach (FD) [15], the spiral tree approach (ST) [2], the stub bundling approach(SB) [14] and the flow motion simulation approach(TNSS) [13]. The results are shown in Figure 11(c), Figure 12, and Table 1. From Figure 11(c), Figure 12, and Table 1, we can see that RFDA-FM has distinctive characteristics and specific advantages.

(1) RFDA-FM can generate flow maps with no acute flow-in angles, no overlaps between nodes and edges, and no edge crosses as ST and SB do. But TNSS and FD have 4 and 1 acute flow-in angles, and FD has 1 overlap between nodes and edges.

(2) RFDA-FM can produce visually smooth and natural edges as TNSS does. The mean value of visual smoothness indexes ($MSI$) for RFDA-FM$_1$, RFDA-FM$_2$, and RFDA-FM$_3$ is 0.751, 0.758, and 0.769 which are a little larger than TNSS and are larger than FD, ST, and SB. The reason is that slightly curved edges with small splitting angles will have large values on the visual smooth index ($SI$), and curve difference necessity ($GC_2$) is considered in our approach. If curviness is the key to be considered, SB may be the best. But as Sun [13] pointed out, the SB layout had an artificial feeling with intentionally curved edges where they could be simply straight. Though the ST layout looks natural, it has the smallest $MSI$ (0.621). The reason is that the splitting nodes in the ST layout are close to the destination nodes and the splitting angles are large [13].

(3) RFDA-FM is better at making nodes not close to the edges, and the distance between nodes and edges can be easily controlled with the parameter $t$. For example, if we set a larger $t$, RFDA-FM$_3$ ($t$=2) will have fewer nodes that are within $100*10^3$m of their nearby edges than RFDA-FM$_1$ ($t$=1) and RFDA-FM$_2$ ($t$=1). By comparing to FD, ST, SB, and TNSS, RFDA-FM does have a larger distance between nodes and edges: RFDA-FM$_2$ has the largest $MDis_{min}$ and RFDA-FM$_3$ has the least nodes which are within $100*10^3$m and $70*10^3$m of their nearby edges.

(4) ST has the smallest total length ($TL$) and RFDA-FM$_2$ has the second smallest $TL$ because the two layouts both have more nodes that are close to their nearby edges. But compared to FD, ST, and SB, RFDA-FM$_2$ has the least nodes which are within $100*10^3$m and $70*10^3$m of their nearby edges. By comparing RFDA-FM$_1$ (a larger $\omega$=0.65) and RFDA-FM$_3$ (a larger $t$=2), the $TL$ of RFDA-FM$_1$ and RFDA-FM$_3$ will increase, but they will have fewer nodes that are close to their nearby edges.

(5) SB has the largest mean value of the user preference index ($MUPI$). The reason may be that SB produces very smooth curves. The $MUPI$ for RFDA-FM$_1$, RFDA-FM$_2$, and RFDA-FM$_3$ are 3.30, 3.04, and 3.43, which means that most of the users moderately prefer the generated layouts by RFDA-FM.

Based on the above analysis, we can see that no single approach is universally better than the others in terms of all the quality criteria in **Section 3.2**. For example, ST has the smallest total length ($TL$), and it also has many nodes which are close to their nearby edges; SB is the most preferred one by users, it also has an artificial feeling; TNSS has visually smooth and natural edges, it also has the largest $TL$. RFDA-FM can well meet all the quality criteria of the flow maps, namely smooth curves, no acute flow-in angles, no overlaps between nodes and edges, no edge crosses, and a moderately preferred tree-like layout with a clear hierarchy of branches as most listed approaches do. Specifically, RFDA-FM has the highest $MSI$ and can be better at keeping destination nodes away from edges.

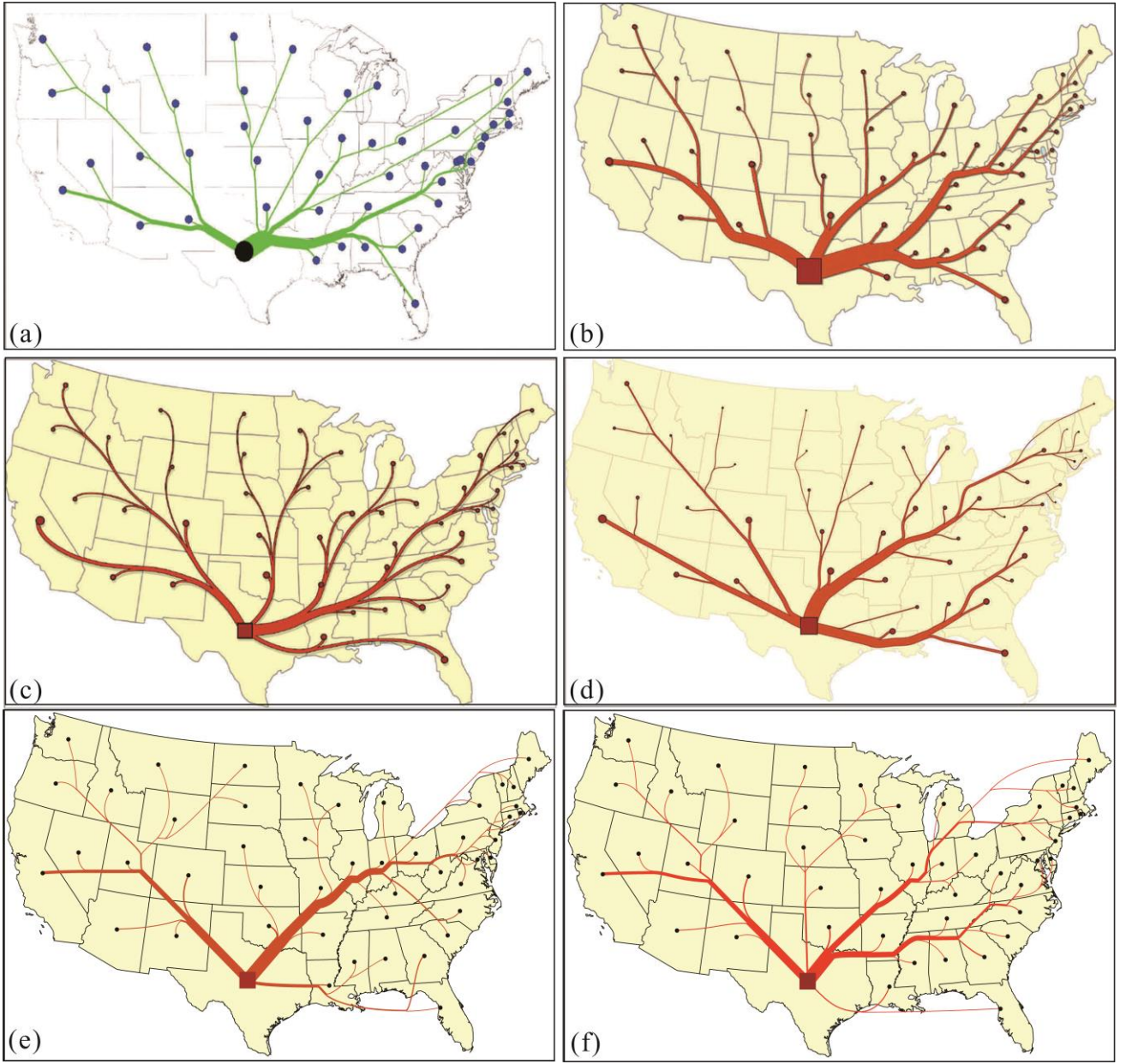

Figure 12. Flow maps for population migration from Texas to other states in the 2000 US Census by existing approaches. (a) The force-directed approach (FD) [15]; (b) The spiral tree approach (ST) [2]; (c) The stub bundling approach (SB) [14]; (d) The flow motion simulation approach (TNSS) [13]; (e) The new approach proposed in this article with $\omega=0.35$, $k=4$, $t=1$ (RFDA-FM$_2$); (f) The new approach proposed in this article with $\omega=0.65$, $k=4$, $t=2$ (RFDA-FM$_3$).

# 6 EXTENSIONS

In the above experiments, mapping spaces are all considered as being homogeneous, while mapping spaces with obstacle areas or heterogeneous mapping spaces are also widely used in practices [2, 13]. As illustrated in **Section 4.2**, our approach can also be applied to these two kinds of mapping spaces.

## 6.1 Case 1. Mapping space with obstacle areas

Population migration from California to other states in the 2000 US Census is taken as experiment data, in which the Great Salt Lake and part of the Mississippi river are taken as obstacle areas. Data source: https://www.ipums.org/; parameter settings are the same as **Section 5.1**.

Results are shown in Figure 13 and Table 2. We observe that a result with no crosses between edges and obstacle areas can be successfully produced if the obstacle areas are considered. Similarly, the total length of the flow map will slightly increase by $0.07*10^6$m (0.26%) due to cross avoidance. While if the obstacle areas are not considered, 2 crosses between edges and obstacle areas will be made, as shown in Area A of Figure 13(a). But the two generated flow maps with our approach can both well avoid node overlaps or edge crosses (except the crosses between edges and important objects), namely, meet the quality criteria in **Section 3.2**.

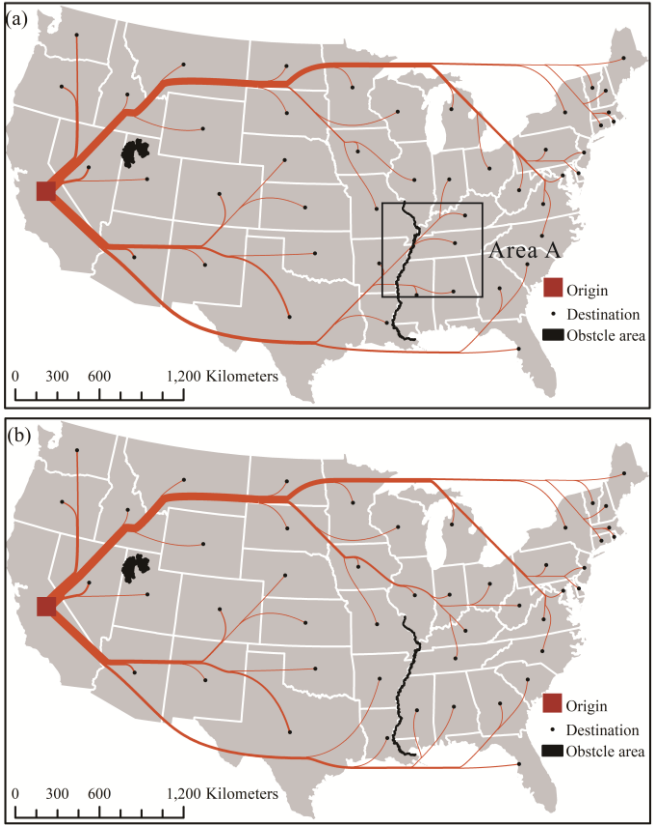

Figure. 13. Flow map for population migration from California to other states in the 2000 US Census. (a) Without consideration of obstacle areas; (b) With consideration of obstacle areas.

### 6.2 Case 2. Heterogeneous mapping space

The good exports from Russia to other European countries ($\geq 0.1\%$) in 2019 are taken as experiment data, in which sea areas need to be avoided as much as possible to save transporting expenses, data source: https://globaledge.msu.edu/countries/russia/tradestats/; parameter settings are the same as **Section 5.1**. To avoid the sea areas as much as possible, the resolution of the grids in sea area ($SR_s$) is set as $SR_s = 1/3 * R_s$ ($R_s$ is defined in Equation (1)). In the maze-solving algorithm, the mapping space is searched grid by grid. If a smaller $R_s$ is set for sea areas, it means more steps are cost by searching across sea areas. The sea areas can then be avoided as much as possible.

From the results in Figure 14 and Table 2, we observe that a result with 5 fewer crosses (as shown in Areas A to E) between edges and sea areas can be successfully produced if the sea areas are considered. As for the 4 generated crosses between edges and sea areas in Figure 14 (b), these crosses are inevitably generated because these involved areas are apart from the main lands. Similarly, the total length of the flow map will slightly increase by $1.43*10^6$m (4.65%) due to sea area avoidance. But the two generated flow maps with our approach can both well avoid node overlaps or edge crosses, namely, meet the quality criteria in **Section 3.2**.

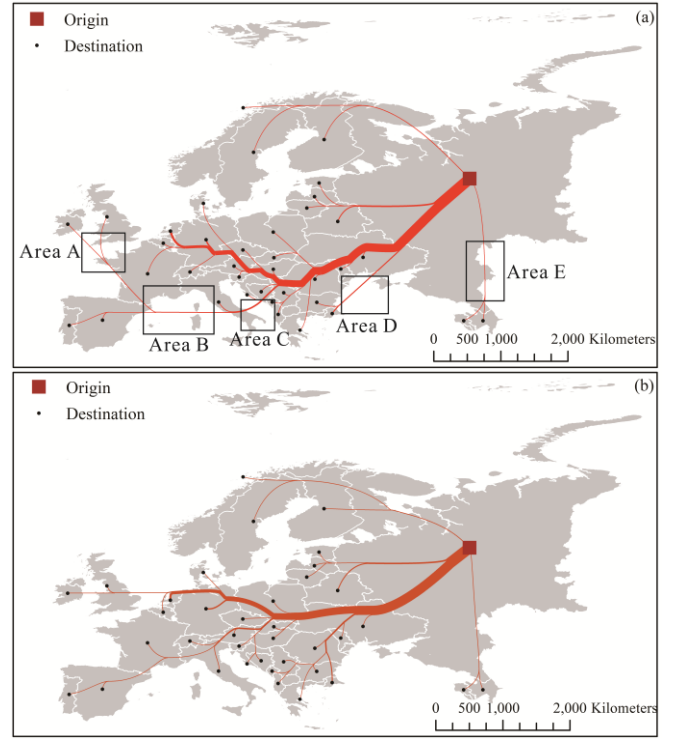

Fig. 14. Flow map for good exports from Russia to other European countries ($\geq 0.1\%$) in 2019. (a) Sea areas are not considered; (b) Sea areas need to be avoided as much as possible.

Table 2. Quantitative assessment on layout quality in different mapping spaces.

| *Measures* | Obstacle areas | | Mapping space | |
|---|---|---|---|---|
| | Without | With | Homogeneous | Heterogeneous |
| $TL$ ($10^6$ m) ↓ | 27.30 | 27.37 | 30.74 | 32.17 |
| $MSI$ ↑ | 0.798 | 0.799 | 0.780 | 0.779 |
| $MDis$ ($10^3$m) | | | | |
| $MDis_{min}$ ↑ | 28.2 | 28.2 | 58.4 | 75.3 |
| $n$ ($MDis$<100) ↓ | 14 | 15 | 7 | 3 |
| $n$ ($MDis$ <70) ↓ | 5 | 5 | 1 | 0 |
| $n$ ($MDis$ <40) ↓ | 1 | 1 | 0 | 0 |
| $n$ ($MDis$ <20) ↓ | 0 | 0 | 0 | 0 |
| $FaN$ (<120°) ↓ | 0 | 0 | 0 | 0 |
| $ECN$ ↓ | 0 | 0 | 0 | 0 |
| $EON$ ↓ | 0 | 0 | 0 | 0 |
| $EIN$ ↓ | 3 | 0 | 4 | 9 |

**Note**: If we consider the obstacle areas, $EIN$ is the number of crosses between edges and obstacle areas; If we consider the heterogeneous mapping space (e.g., sea areas need to be avoided as much as possible), $EIN$ is the number of crosses between edges and sea areas.

## 7 DISCUSSIONS
### 7.1 Strategy effectiveness analysis

To generate a satisfactory flow map layout, some strategies are applied in our approach and are summarized in Table 3. To validate the feasibility of these strategies, ablation experiments are performed. The baseline is the result for evaluation in **Section 5.2**, Figure 11(c). Three layouts without the strategies in Table 3 are then produced, as in

Figure 15. The statistical results are shown in Table 4.

Table 3. Strategies applied in our approach.

| Name | Descriptions | Location |
|---|---|---|
| $St_1$ | Path length refinement by considering the acute flow-in angle avoidance ($RC_1$). | Section 4.3.1.2 |
| $St_2$ | Searching range definition by considering overlaps avoidance and suitable distance maintenance between nodes and edges ($RC_3$, $RC_5$) | Section 4.3.2.3 |
| $St_3$ | Path importance definition in which the flow path straightly connects to the origin is preferred, which helps minimize the total length | Section 4.3.3 |

(1) $St_1$ is applied to avoid acute flow-in angles. As shown in Figure 15(a) and Table 4, if $St_1$ is not applied, 2 acute flow-in angles will be made. But if $St_1$ is applied, the flow-in angles can be effectively avoided, as in Figure 11(c); But longer branches may be required to avoid acute flow-in angles, and the total length will slightly increase by $0.02*10^6$m (0.08%). Nevertheless, the layouts which are generated with or without $St_1$ can all well meet the quality criteria in **Section 3.2** (except the acute flow-in angles avoidance). Thus, we can conclude that $St_1$ is effective to avoid acute flow-in angles, but it will also slightly increase the total length.

(2) $St_2$ is applied to avoid overlaps and maintain suitable distances between nodes and edges. As shown in Figure 15(b) and table 4, if $St_2$ is not applied, 7 overlaps between nodes and edges will be made, and 10 nodes with their distance to nearby edges less than $40*10^3$m. If $St_2$ is applied, the overlaps between nodes and edges will be effectively avoided, Figure 11(c); and no nodes are with a distance to nearby edges less than $40*10^3$m; but longer edges may be required to avoid overlaps and maintain suitable distance between nodes and edges, the total length will increase $0.70*10^6$m (2.84%). These prove that $St_2$ can effectively avoid overlaps and maintain suitable distance between nodes and edges, and the total length may also increase if $St_2$ is applied.

(3) $St_3$ is applied to minimize the total length of the flow map. By comparing Figure 11(c) and Figure 15(c), we can see that the total length will increase $1.19*10^6$m (4.83%) if $St_3$ is not applied. Nevertheless, the layouts which are generated with or without $St_3$ can both well meet the quality criteria in **Section 3.2**. These prove that $St_3$ is very helpful to reduce the total length of a flow map.

Table 4. Quantitative assessment on layout quality for strategy effectiveness analysis.

| Measures | $St_1$ | $St_2$ | $St_3$ |
|---|---|---|---|
| TL ($10^6$ m) ↓ | 24.61 | 23.93 | 25.82 |
| MSI ↑ | 0.757 | 0.772 | 0.767 |
| MDis ($10^3$m) | | | |
| $MDis_{min}$ ↑ | 52.5 | 0 | 52.5 |
| n (MDis<100) ↓ | 15 | 16 | 13 |
| n (MDis <70) ↓ | 8 | 14 | 6 |
| n (MDis <40) ↓ | 0 | 10 | 0 |
| n (MDis <20) ↓ | 0 | 7 | 0 |
| FaC (<120°) ↓ | 2 | 0 | 0 |
| EEC ↓ | 0 | 0 | 0 |
| EOC ↓ | 0 | 7 | 0 |

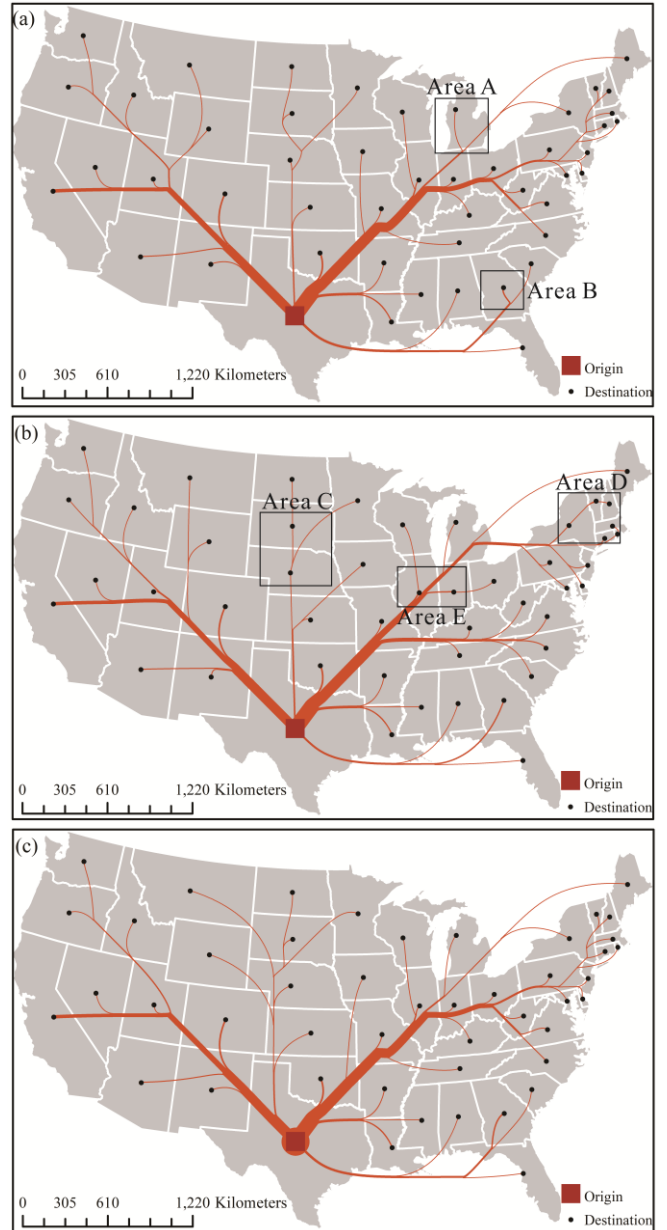

Figure 15. Flow maps for population migration from Texas to other states in the 2000 US Census which are produced without (a)$St_1$, (b)$St_2$, and (c) $St_3$.

## 7.2 Parameter sensitivity analysis

In this section, we analyze the influence of the parameter settings on the layouts and show how these parameters can intuitively control the layouts.

(1) $\omega$ in Equation (3)

$\omega$ is used to define the path length by considering the total length minimization ($RC_3$). Smaller $\omega$ means smaller total length (TL). We set $\omega=0.65$ in our experiment for evaluation and the result is shown in Figure 11(c). If a smaller $\omega$ is set ($\omega=0.35$), the paths from the destination nodes tend to flow in nearby paths, and the result is shown in Figure 12(e). If a larger $\omega$ is set ($\omega=1.0$), the paths from the destination nodes tend to flow in a location that is near the origin node, and the result is shown in Figure 16. The TL will increase with the increase of $\omega$: the TL is $22.58*10^6$m if $\omega=0.35$, $24.63*10^6$m if $\omega=0.65$, and $41.95*10^6$m if $\omega=1.0$.

Similarly, the generated layouts will have longer branches with the increase of $\omega$, as shown in Figures 11(c), 12(e), and 16. Furthermore, the nodes which are with small distance to their nearby edges will both increase with smaller or larger $\omega$, and the $n(MDis<100)$ is 17 if $\omega=0.35$, 25 if $\omega=1.0$, but 14 if $\omega=0.65$. Nevertheless, the generated layouts with different $\omega$ can all well meet the quality criteria in **Section 3.2**, e.g., no overlaps between nodes and edges and no edge crosses. If the users prefer a layout with a smaller total length, a smaller $\omega$ is recommended; if the users prefer a layout with longer branches, a larger $\omega$ is recommended.

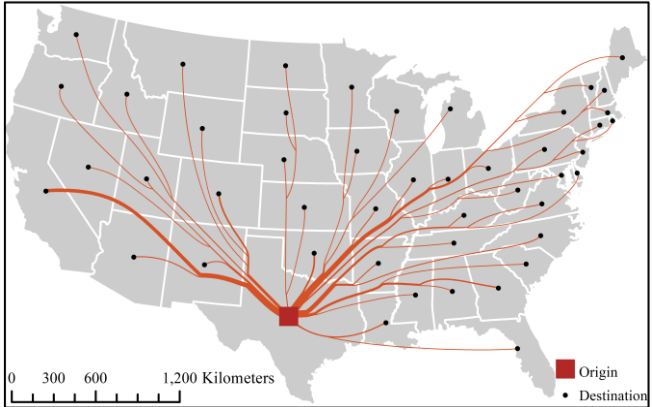

Figure 16. Flow maps for population migration from Texas to other states in the 2000 US Census with $\omega=1.0$.

(2) $k$ in Equation (6)

$k$ defines the $k$-order surrounding of the grid and is used for potential flow accumulation ($Pf$) computation. The direction to a grid with a higher $Pf$ is prior to be searched in the maze-solving algorithm, which helps minimize the total length ($TL$). We set $k=1$ in our experiment for evaluation and the result is shown in Figure 11(c). The flow maps with $k=0$ (smaller $k$) or $k=8$ (larger $k$) are shown in Figure 17. As shown in Figures 11(c) and 17, we can see that the $TL$ will both increase with smaller or larger $k$: the $TL$ is $24.89*10^6$m if $k=0$ and $24.74*10^6$m if $k=8$, and the $TL$ is $24.60*10^6$m if $k=4$. Thus, a suitable $k$ needs to be set in practice and $k=4$ is recommended. Nevertheless, the generated layouts with different $\omega$ can all well meet the quality criteria in **Section 3.2**.

(3) $t$ for searching range definition considering overlap avoidance and suitable distance maintenance between nodes and edges ($RC_3$ and $RC_5$)

$t$ is used to avoid overlaps and maintain suitable distance between nodes and edges ($RC_3$ and $RC_5$), and we set $t=1$ in our experiment for evaluation and the result is shown in Figure 11(c). If a smaller $t$ is set ($t=0$), it means overlap avoidance is not considered, and the result is shown in Figure 15(b). Then 7 overlaps between nodes and edges will be made, and 10 nodes with their distance to nearby edges less than $40*10^3$m. If a larger $t$ is set ($t=2$), the result is shown in Figure 12(f). Then the number of nodes that are within a small distance ($<100*10^3$m) of their nearby edges will reduce, as 6 if $t=2$ and 15 if $t=1$. But if $t>0$, no overlaps between nodes and edges will be made. The users can set a suitable $t$ ($t>0$) in practice according to their demands. If the users prefer a larger distance between nodes and edges, a larger $t$ is recommended.

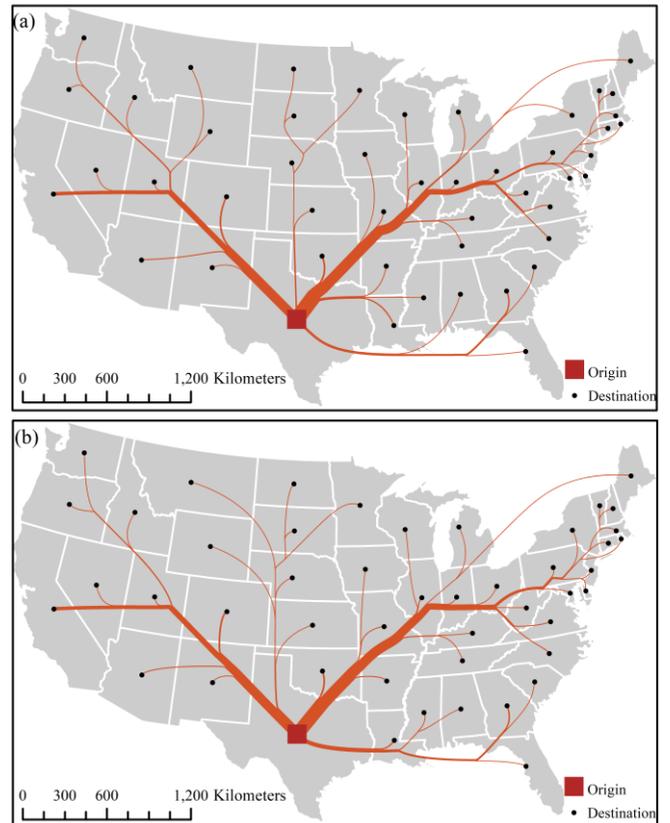

Figure 17. Flow maps for population migration from Texas to other states in 2000 US Census. (a) $k=0$; (b) $k=8$.

(4) Other parameters

Some other parameters may also need to be set in the proposed approach e.g., the resolution ($R_s$) of the modeling DEM data, and the default priority of searching directions in the maze-solving algorithm. These parameters are set as default according to previous works. For example, large $R_s$ will lead to an error because the closest points may fall into the same grid; while a small $R_s$ may increase the computational cost. $R_s$ is set as Equation (1) according to Hengl [43]. These parameters can also be set according to user demands.

### 7.3 Limitation analysis

Although suitable flow maps in different kinds of mapping spaces can be generated by using the proposed approach. It also has some limitations.

(1) The mapping space is modeled as a flat surface with the digital elevation model (DEM) and the grids in DEM data are assigned different types. But the elevation value is another key property in DEM data and is not considered in our approach. More complex flow data can be modeled with the DEMs if the elevation values are also considered. To improve the generalization ability of our approach, the elevation values need to be considered in our future works.

(2) As the flow paths are generated with an iterative process by a maze-solving algorithm. Thus, a result that meets the quality criteria as much as possible is just obtained each time, but not an optimal one. The result can be improved in our approach by introducing some other strategies, such as a backtracking strategy if unexpected results occur.

(3) Though the heterogeneous mapping space is considered in our approach, nodes of the mapping data may also distribute heterogeneously. For example, if the *k*-order surrounding definition for potential flow accumulation (Section 4.3.2.2) is adapted to different densities, different layouts may be produced.

(4) The proposed approach focuses on mapping data from one origin to many destinations. But in practice, data with multiple origins to many destinations or many origins to many destinations are also commonly used data, our approach needs to be extended for these data in future works.

(5) The generation of the flow map for the experiment data with 1 origin and 46 destinations will cost 6 minutes with the proposed approach. Though an approximate approach is also provided to reduce the search space in our uploaded code which can shorten the time to within 1 minute. More efficient strategies to speed up the proposed approach may need to be provided. Many efficient strategies have been introduced in digital elevation model data processing, and these strategies can also be applied to our approach.

## 8 CONCLUSIONS

In this paper, to generate a one-to-many flow map for visualization of movement data, we propose a river flow directions assignment algorithm over flat surfaces in digital elevation models (DEM) by modeling the mapping space as DEM data. Experiments indicate that the flow maps obtained by the proposed approach can achieve a higher quality in keeping nodes away from edges without node overlaps or edge crosses. Furthermore, the experiments demonstrate that our approach is also applicable to heterogeneous mapping space and mapping space with obstacle areas. Besides, the quality criteria can be intuitively controlled by setting parameters for users.

Future works will focus on: (1) Adaptive algorithm for different distributions of mapping data; (2) Visualization of flow map with multiple origins and dynamic data; (3) Improvement of algorithm efficiency.

## ACKNOWLEDGMENTS

The authors wish to thank Tingzhong Huang for his help in data collection and Yalong Yang for sharing his JavaScript code. This work was supported in part by a grant from The National Natural Science Foundation of China (No.41871378) and The Research Development Fund of Zhejiang A & F University (2020FR083).

**ZHIWEI WEI** received a Ph.D. degree in cartography and geographic information science from Wuhan University in 2020. He is currently an assistant professor at the Aerospace Information Research Institute, Chinese Academy of Sciences, Beijing, China. His research interests include automatic map generalization, automatic map design, and spatial data visualization.

**SU DING** received a Ph.D. degree in geography from Wuhan University in 2020. She is currently an assistant professor at the College of Environmental and Resource Sciences, Zhejiang A & F University. Her research interests include geospatial information analysis and modeling, and spatial data mining.

**WENJIA XU** received a Ph.D. degree in signal and information processing from the University of Chinese Academy of Sciences in 2021. She also worked as a visiting Ph.D. student at the Max-Planck Institute for Informatics in Saarbrucken, Germany between 2019-2020. She is currently a research associate professor at the School of Information and Communication Engineering, Beijing University of Posts and Telecommunications. Her research interest includes learning with limited supervision for computer vision tasks, explainable machine learning, and data visualization.

**YUANBEN ZHANG** received a B.S. degree from the University of Science and Technology of China in 2011, and an M.S. degree from the Institute of Electronics, Chinese Academy of Science, Beijing, China in 2014. He is currently an assistant professor at the Aerospace Information Research Institute, Chinese Academy of Sciences, Beijing, China. His research interests include geospatial data mining and visualization analysis.

**YANG WANG** received a Ph.D. degree in graphics from Peking university in 2014. She is currently an associate professor and a master advisor at the Aerospace Information Research Institute, Chinese Academy of Sciences, Beijing, China. Her research interests include geospatial information analysis and visualization.